# Parameters of the Menzerath-Altmann law:
# Statistical mechanical interpretation as applied to a linguistic organization

Sertac Eroglu

*Department of Physics, Eskisehir Osmangazi University, Meselik, 26480 Eskisehir, Turkey*
*e-mail: seroglu@ogu.edu.tr*

**ABSTRACT**

The distribution behavior dictated by the Menzerath-Altmann (MA) law is frequently encountered in linguistic and natural organizations at various structural levels. The mathematical form of this empirical law comprises three fitting parameters whose values tend to be elusive, especially in inter-organizational studies. To allow interpretation of these parameters and better understand such distribution behavior, we present a statistical mechanical approach based on an analogy between the classical particles of a statistical mechanical organization and the number of distinct words in a textual organization. With this derivation, we achieve a transformed (generalized) form of the MA model, termed the statistical mechanical Menzerath-Altmann (SMMA) model. This novel transformed model consists of four parameters, one of which is a structure-dependent input parameter, and three of which are free-fitting parameters. Using distinct word data sets from two text corpora, we verified that the SMMA model describes the same distribution as the MA model. We propose that the additional structure-dependent parameter of the SMMA model converts the three fitting parameters into structure-independent parameters. Moreover, the parameters of the SMMA model are associated with a corresponding physical interpretation that can lead to characterization of an organization's thermodynamic properties. We also propose that many organizations presenting MA law behavior, whether linguistic or not, can be examined by the SMMA distribution model through the properly defined structural degeneracy parameter and the energy associated states.

1. **Introduction**

The Menzerath-Altmann (MA) law is one of the well-known stochastic laws in quantitative linguistics and has been considerably put into practice. The law principally states that 'the longer the construct the shorter its constituents' [1]. This general statement has been employed to measure regularities in the structural organization of many natural languages at various organizational levels, including phonemic, morphemic, syntactic, and textual [2-5]. Furthermore, the extent of this distribution behavior is not limited to linguistic organizations, in fact, the MA law has also been shown to describe some quantifiable regularity in a variety of semiotic and biologic organizations as well [6-12].

The familiar mathematical definition of the MA law which describes the dependence of a construct's size (or length) $y$ on its constituent's size (or length) $x$ is given by [13]

$$y(x) = A x^b e^{-cx}. \tag{1}$$

Here $A$, $b$ and $c$ are the model parameters. Eq. (1), power-law with exponential cutoff, should be viewed as a continuous distribution function of the structural organization under investigation, and the empirical parameters are uniquely determined for the best fit. Linguistic or not many organizational constructs mostly comprise discrete constituent sizes, $x_i$, while it cannot be said the same for construct size (in some cases a mean size), $y$, which is not priory discrete but not truly continuous.

Using the MA law for the detection of regularities at the word length level has primarily been the attention of the correlation studies between the length of the words occurring in a text and the length frequency of each word's constituents [14,16]. On the other hand, our recent study [17] reports that the length distribution of vocabulary[1] or distinct words (DWs) in a large text obeys the MA law distribution behavior. This distribution behavior can be translated as 'the number of relatively short length DWs in a text increases when the text length increases'. The study specifically reveals that the MA law, a special case of gamma distribution function, is quite accurate in describing the DW length distribution, in letter count, in a large text. Hence, the MA law is more descriptive model than the familiar Heap's law [18]; i.e., 'the number of DWs increases with increasing text length'.

Although the MA law is a well-recognized distribution model in the study of linguistics and naturally occurring phenomena that present language-like organizational behavior, yet, there is no convincing statistical support for the law's widespread validity and the substantiated interpretation of its parameters. There have been numerous attempts to elucidate the model parameters from a linguistics point of view by using comparative parameter analysis [19-20]; however, the interpretation of the parameters is still controversial. This fact hampers research on two levels; firstly, the ability to reach decisive conclusions during comparative studies, i.e. between different organizational levels or between different source of organizations (e.g., languages), and secondly, the ability to reach a comprehensive understanding of organizational dynamics. In turn, these drawbacks prevent realization of the full potential of the MA law.

In an effort to estimate some thermodynamic properties of a linguistic organization, several studies have proposed to implement statistical mechanics tools to uncover the fundamental regularities in linguistic organizations. Some of these studies [21, 23] simply made use of the universality of Maxwell-Boltzmann's exponential term, $\exp(-E/k_B T)$. Some other studies [24, 25], on the other hand, have approached the problem from information content perspective by utilizing Shannon entropy, $H = -\sum_j p_j \log p_j$. All of these studies provide viable information for a given organization; however, due to varying structural properties it becomes intricate to relate the obtained thermodynamic properties of dissimilar structural organizations.

---

[1] Henceforth, we will refer to vocabulary of a text as distinct words (DWs) for convenience. In general, DWs of a text can be considered as the set of dissimilar constituents of a construct at the level of organization under investigation.

In this study, our objective is to present a theoretical framework for the derivation of the empirical MA distribution model from a statistical mechanical perspective. The derived model was referred to as the statistical mechanical MA (SMMA) law. The study revealed that the constituent distribution of a linguistic organization which is presenting the MA law behavior can alternatively be described by the SMMA law. We showed the MA and SMMA laws are actually the same distribution functions having different sets of parameters. The derivation of the SMMA law was based on the analogy between the non-interacting classical particles of a statistical mechanical organization and the DWs of a textual organization. Since the SMMA law was derived using the description of the structural organization at the microscopic (constituent) level to obtain the organizational properties at the macroscopic (construct) level, the procedure establishes a firm foundation for interpreting the derived model parameters in terms of physical concepts. We finally proposed that if the structural organization under investigation (which could be from various disciplines) presents the MA law behavior, this same procedure can be implemented to characterize constituent diversity dynamics of that structural organization in terms of thermodynamic concepts.

The paper is organized as follows: Section 2 describes the derivation process of the MA law using statistical mechanical concepts and tools. For the sake of consistency in notation, Section 2 includes a brief statistical mechanical description of the accessible states of a classical particle system, and it is followed by the analogous treatment of the accessible states of DWs in a text. The assessment of the SMMA law, physical interpretation of its parameters and a demonstrating study are presented in Section 3. Section 4 provides the concluding remarks of this study and the extent of the obtained results.

## 2. Physical analogy and the model derivation

*2.1. Accessible microstates of a classical particle organization*

We start the derivation of the SMMA model by introducing a brief review of the familiar physical system of classical, or Maxwell-Boltzmann, particles. Suppose that the total energy of the system is $E$, and the system contains the total of $N$ non-interacting particles, e.g., atoms, molecules or elementary particles. Furthermore, the particles are distinguishable and they are distributed over a set of quantized energy levels $\varepsilon_1, \varepsilon_2,\ldots,\varepsilon_k$ such that the energy of a particle at the $i^{\text{th}}$ energy state is $\varepsilon_i$. Each energy state has an associated degeneracy $g_1, g_2,\ldots,g_k$ with a corresponding number of occupations $n_1, n_2,\ldots,n_k$. There are no restrictions as to the number of particles in any given state. The two requirements that are imposed on the number of occupations distributed over the energy states are as follows:

a) The total number of particles, $N$, is fixed:

$$N = n_1 + n_2 + \ldots + n_k = \sum_i n_i . \qquad (2)$$

b) The total energy, $E$, of the system is constant:

$$E = \varepsilon_1 n_1 + \varepsilon_2 n_2 + \ldots + \varepsilon_k n_k = \sum_i \varepsilon_i n_i . \qquad (3)$$

The average energy of this classical particle organization is

$$\bar{E} = \sum_i \varepsilon_i \, p_i(\varepsilon_i), \qquad (4)$$

where $p_i(\varepsilon_i)$ is the probability of a particle being in the energy state $\varepsilon_i$. The probability distribution of the particles is explored using combinatorial statistical mechanics analysis. The multinomial coefficient $W$ that is the total number of ways in which $N$ distinguishable particles displaying a particular set of distribution $\{n_i\}$ is defined by

$$W(\{n_i\}) = \frac{N!}{n_1! n_2! \ldots n_k!} = \frac{N!}{\prod_i n_i!}. \qquad (5)$$

The disorder number $\Omega$ is defined as the number of microstates available to a macrostate, or number of accessible microstates. If the energy states are not degenerate, the disorder number is equal to $W$. However, in general, the energy states contain associated degenericies, degenerate states are the states with the same energy, and then the disorder number is given by

$$\Omega(\{n_i\}) = W \prod_i g_i^{n_i} = N! \prod_i \frac{g_i^{n_i}}{n_i!}. \qquad (6)$$

This is the general equation for the accessible microstates of a classical particle system. In the following section we aim to define the analogous accessible microstates definition for the DWs in a text.

*2.2. Accessible microstates of a DW organization*

A text exhibits several levels of structural, syntactic and semantic organization. We have to emphasize that the presented procedure is applicable to many linguistic or nonlinguistic organizational levels which present the MA law behavior. However, due to the concerns such as straightforward presentation of the derivation and assessment of the theoretical result with readily available data; in this study, the textual organization level of interest is the simple structural letter-string organization of DWs in a text.

Consider that we examine a large text, or corpus, to increase the number and variety of samples for statistical completeness, and suppose the corpus contains a total number of $N_T$ words, $N$ of which are distinct, i.e. distinguishable. According to the linguistic rules and the desired information content to be transmitted, a particular text's words interact with each other in the word-string organization. If we consider the text at the level of DW organization, however, there is no explicit association or regularity of DWs occurrence. This condition suggests that the DWs of a text present non-interacting behavior. Therefore, in our suggested physical analogy, a corpus consisting of $N$ number of DWs can be treated as a classical particle system of $N$ particles, in which each DW corresponds to a non-interacting and distinguishable particle of the system.

The next step in treating a corpus as a classical particle system is to define the DW organization's energy states by equating word length with word energy; i.e., the length of each word on the basis of letter count equals the word's energy (or effort). Empirical observations have shown that human behavior, including articulation, tends to obey the principle of least effort, for a straightforward reason: The longer the word the longer the time it takes to read, write and perceive that word. The principle of least effort was originally proposed by Zipf [26], and recently a more direct connection between energy and information theory has been explored [27]. Accordingly, the energy-preserving preference of language users supports our word length and word energy analogy as a quite realistic assumption during written or verbal communication. Thus, we will use the terminologies 'word length, in letter count,' and 'word energy' interchangeably, for the rest of the paper.

As a result, the DW energies are distributed over a set of quantized energy levels given by quantized length states $l_1, l_2,\ldots,l_k$ in letter count such that the energy of a DW at the $i^{th}$ energy state is simply the word's letter count $l_i$. Each DW energy state has an associated degeneracy $g_1, g_2,\ldots,g_k$ with a corresponding number of occupations $n_1, n_2,\ldots,n_k$. There are no restrictions as to the number of DWs in any given state. As in the classical particle organization, the two requirements that are imposed on the number of occupations distributed over the DW length states are as follows:

a) The total number of DWs, $N$, is fixed:

$$N = n_1 + n_2 + \ldots + n_k = \sum_i n_i,  \quad (7)$$

where $n_i$ is the number of DWs, distinguishable words, in the $i^{th}$ state; i.e., the number of DWs having the same length $l_i$ in letter count.

b) Note that we only consider the word length distribution of DWs in a text, so the frequency and the occurrence positions of the words are insignificant. Similar to the total energy in the classical particles system, Eq. (3), the total word length count of DWs $L$ is finite:

$$L = l_1 n_1 + l_2 n_2 \ldots + l_k n_k = \sum_i l_i n_i. \quad (8)$$

The average length of this DW organization is

$$\bar{L} = \sum_i l_i \, p_i(l_i), \tag{9}$$

where $p_i(l_i)$ is the probability of a DW's being in the length state $l_i$. The rest of the accessible states derivation of the DW organization is the same as in the classical particle organization, Eq. (6).

The key at this stage of the derivation procedure is to define the degeneracies associated with the length states of DWs. Since the accessible number of DWs at a particular length state is called the degeneracy of that state, the degeneracy of the $i^{\text{th}}$ word length state is theoretically equal to $\omega^{l_i}$. Here $\omega$ is the number of letters in the alphabet of the language in which the text is written. In general terms, $\omega$ can be defined as '*structural degeneracy parameter*'. $\omega$ is principally equal to the total number of distinct units, i.e. letters, from which the DWs can be formed. For instance, in English the degeneracy of the three-letter-long DW state, $l_3$, is $26^3 = 17576$; i.e., there are 17576 possible manifestations of a three-letter-long DW. In practice, however, the occupancy of the accessible degenerate states for a given word length state is primarily governed by two counteracting effects: (i) the linguistic rules and restrictions prohibit the generation of some of the accessible degenerate states, DWs. For instance, as long as they are not abbreviations, English vocabulary does not consist of some of the words in $l_3$ length state such as '*aps*', '*eps*', '*iop*', and many more. This effect has a relatively more significant impact on shorter length DWs than longer length DWs; in turn, the DW length is forced to have higher word length values for the generation of new DWs, and (ii) the principle of least effort effect, on the other hand, forces the words to have shorter lengths for feasible articulation, as discussed earlier. At equilibrium, these two counteracting effects define an optimum word length value of which the occupancy tendency of the degenerate states will be relatively higher around that particular word length value.

The primary proposition of statistical mechanics is that all the states of a physical system are equally likely accessible. This assumption holds for isolated systems; however, in many non-isolated systems, a certain state's occurrence can be more probable than that of others. The aforementioned two counteracting effects cause the DW length organization in a text to behave as a non-isolated physical system; i.e., length states present favorability. Therefore, to account for the length favorability of DW states, we implement an *ansatz* such that a particular outcome's probability $p_i$, which is associated with the occurrence of $i^{th}$ outcome, is weighted by means of positive-valued power $\alpha$ of the discrete variable $l_i$. This weighted probability requires a weighted number of degeneracy at the $i^{th}$ state, which can be defined as

$$g_i = \left(\omega^{l_i} l_i^{\alpha}\right). \tag{10}$$

If the degeneracy of the $i^{th}$ state is not biased by the mechanism of the aforementioned two counteracting effects, i.e., $\alpha$ is equal to 0, the number of degeneracy of the $i^{th}$ state in Eq. (10) reduces to its equally accessible states form, $\omega^{l_i}$, as expected. Finally, the number of ways in which $N$ distinguishable words of a text can take place is obtained by substituting Eq. (10) into Eq. (6)

$$\Omega(\{n_i\}) = N! \prod_i \frac{\left(\omega^{l_i} l_i^{\alpha}\right)^{n_i}}{n_i!}. \tag{11}$$

*2.3. Derivation of the DW distribution model*

The most probable distribution is determined by the realization of the set of occupations $\{n_i\}$ that maximizes accessible microstates, $\Omega$. Note that $\Omega$ is defined on a subset of the real valued numbers and satisfies the condition of $\Omega(x) \leq \Omega(y)$ for all $x \leq y$; i.e., $\Omega$ is a monotonically increasing function. Hence, maximizing $\Omega$ is the same as maximizing $\ln \Omega$. This functional behavior allows us to easily approximate the factorial terms in Eq. (11). Next, assume each $n_i$ is sufficiently large, which implies that $N$ is very large, ideally in the limit as $N \to \infty$; in this case Stirling's approximation, $\ln(n_i!) \approx n_i \ln(n_i) - n_i$ provides a quite accurate estimate for the factorial terms in Eq. (11) and we obtain

$$\ln \Omega = \left[ N \ln N - N + \sum_i \left( \ln g_i^{n_i} - n_i \ln n_i + n_i \right) \right]. \tag{12}$$

$\ln \Omega$ is maximized for $n_i$ value, which satisfies the following condition:

$$d \ln \Omega = -\sum_i \left[ \ln n_i - \ln \left( \omega^{l_i} l_i^{\alpha} \right) \right] dn_i = 0. \tag{13}$$

Since $dn_i$'s are related to each other by the constraints given in Eqs. (7) and (8), the solution to this extreme value problem is achieved by scrutinizing the constraints associated with the Lagrange multipliers $\phi$ and $\theta$, i.e., the well-known Lagrange multipliers' method,

$$\phi \sum_i dn_i = 0, \tag{14}$$

and

$$-\theta \sum_i l_i \, dn_i = 0. \tag{15}$$

In Eq. (15), the minus sign is arbitrary; however, for positive valued $l_i$ the sum appropriately converges. By substituting Eq. (14) and Eq. (15) into Eq. (13), we get

$$\sum_i \left[ \ln n_i - \ln\left(\omega^{l_i} l_i^{\alpha}\right) - \phi + \theta l_i \right] dn_i = 0. \tag{16}$$

Now we assume all d$n_i$'s are independent of each other by suitably chosen $\phi$ and $\theta$ values that fulfill the conditions required for the constraints, Eq. (7) and Eq. (8). The condition for Eq. (16) is satisfied when the term inside the square bracket identically vanishes for each $i^{th}$ state such that

$$\ln n_i = \ln\left(\omega^{l_i} l_i^{\alpha}\right) + \phi - \theta l_i \tag{17}$$

which leads to the most probable length distribution of the DWs as follows

$$n_i(l_i) = \omega^{l_i} e^{\phi} l_i^{\alpha} e^{-\theta l_i}. \tag{18}$$

Using the constraint in Eq. (7), Eq. (18) can be alternatively rewritten as

$$n_i(l_i) = N \frac{\omega^{l_i} l_i^{\alpha} e^{-\theta l_i}}{Z}, \tag{19}$$

where $Z$ is the partition function, and it is defined by

$$Z = \sum_i \omega^{l_i} l_i^{\alpha} e^{-\theta l_i}. \tag{20}$$

The correspondence between the models and the physical implications of the parameters for the DW length distribution in a large text are presented in the following section.

## 3. Results and Discussion

### 3.1. Parameters of the models

Equation (18) is the four-parameter SMMA model that was derived to characterize the DW organization in a large text. One of the parameters, structural degeneracy parameter $\omega$, in the SMMA model is not a free parameter, it is a fixed or structure-specific parameter. The remaining three free parameters are to be determined experimentally. In Eq. (18), let

$$\left.\begin{array}{l} \alpha = b \\ e^{\phi} = A \\ \theta = c + \ln(\omega) \end{array}\right\} \tag{21}$$

then, the SMMA model reduces to the discrete form of the MA model, see Eq. (1),

$$n_i(l_i) = A l_i^b e^{-c l_i}. \tag{22}$$

This suggests that both the SMMA model, Eq. (18), and its reduced form of the MA model, Eq. (22), theoretically describe the very same distribution function with two different sets of parameters which are related to each other given by Eq. (21). As a result, both models are identical in their functional behavior, i.e.,

$$n_i(l_i) = A l_i^b e^{-c l_i} \equiv e^{\phi} l_i^{\alpha} \omega^{l_i} e^{-\theta l_i}. \tag{23}$$

In this study, the particular organization that we are examining is the length distribution, in letter count, of DWs (constituents) in a large text (construct). Our recent study [17] demonstrated the validity of the MA model in describing the DW distribution for two corpora written in different languages, the Brown Corpus (English) and the METU Corpus (Turkish). Since the related question is whether the derived SMMA model experimentally predicts the same DW distribution as the MA model does, we utilized the same data set in Ref. [17] for consistency, as seen Table 1.

> Table 1 is placed about here

The discrete data were fitted to the MA model and to the SMMA model with the best fitting parameter sets given in Table 2. Since there are 26 letters in the English alphabet and 29 letters in the Turkish alphabet, the structural degeneracy parameter in the SMMA model was taken to be $\omega=26$ for the Brown Corpus and $\omega=29$ for the METU Corpus. The non-linear regression analysis was performed by using Levenberg-Marquardt algorithm [28], also known as the damped least-square fitting method. The algorithm initially starts with the user defined parameter guesses, and iteratively generates slight variations in the parameter values. At each iteration, the sum of the squared error between the observed data and the predicted fit, chi-square value, is calculated and the best fit is found by minimizing the chi-square value. One might improve the goodness of the fit by utilizing different regression analysis method; in this study, however, our priority was to present the correspondence between the MA and SMMA models by applying the same regression analysis method to both models.

> Table 2 is placed about here

Table 1 shows the predicted DW distribution values for both corpora using both distribution models with the parameter sets given in Table 2. The results indicated that the DW distribution values for both models are the same with negligible differences in some states' predictions. Another evidence supporting our claim that both distribution models experimentally perform in the same manner is the identical values of $R$, linear correlation coefficient, and $R^2$, coefficient of determination, for both models as seen in Table 2. To graphically illustrate the correspondence between the models, the observed data and the predicted distribution curves by the MA model, Eq. (22), and the SMMA model, Eq. (18), are shown in Fig. 1(a) and Fig. 1(b) for the Brown Corpus and for the METU Corpus, respectively. Notice that both predicted DW distribution curves exactly overlap for both corpora. In conclusion, these experimental indications confirmed that the SMMA and the MA laws are the same distribution models, and the SMMA model is the transformed (generalized) form of the MA model.

Figure 1 is placed about here

As theoretically proposed in Eq. (21), the experimental values of the parameters $b$ and $\alpha$ are essentially equal to each other and independent from the model utilized, see Table 2, which suggests that $b$, or $\alpha$, is just responsible for predicting the height of the distribution maxima. This behavior can be confirmed by simulating the varying $\alpha$ values. While $\alpha$ parameter is invariant under the model transformation, $\alpha \cong b$, and $e^{\phi} \cong A$; the values of parameter $c$ of the MA model and its corresponding parameter $\theta$ in the SMMA model were observed to be different (Table 2). Moreover, the values of parameters $c$ and $\theta$ can be confirmed to be related to each other as dictated by the structural degeneracy parameter given by Eq. (21). This result revealed that the structure dependent information of the organization is implicitly embedded into the MA model, but explicitly included in the SMMA model. In other words, the $\omega^{l_i}$ term in the SMMA model inputs organization characteristic information to the distribution function and, in turn, converts the other model parameter, $c$, into the structure-independent model parameter, $\theta$. Thus, the obtained $\theta$ parameter values are independent from the structural differences for the organizations under investigation. This is an exceedingly useful characteristic property of the SMMA model, especially in the comparative studies of different organizations, as demonstrated in the following sections.

*3.2. Physical interpretation of the SMMA model parameters*

Since the SMMA model, Eq. (18), was derived by using statistical mechanical concepts and tools, the model parameters have the following noticeable physical interpretations: The exponential term, $\exp(-\theta l_i)$, is analogous to the Maxwell-Boltzmann exponential term, $\exp(-\varepsilon_i/k_B T)$, in a classical particle distribution. Hence, the parameter $\theta$ is equivalent to $\beta$ which is equal to the reciprocal of the fundamental physical energy unit $k_B T$; i.e., $\beta = 1/k_B T$. Here $k_B$ is Boltzmann's constant, T is absolute temperature, and $k_B T$ is the energy associated with each microscopic degree of freedom. For a given classical particle system, the average kinetic energy associated with a particle's degree of freedom (e.g., the translational motion) is given by $k_B T$. Therefore, the mean translational kinetic energy per particle is proportional to the temperature, and the multiplication of this average energy by the total number of particles is simply equal to the thermal energy of the system. When absolute temperature drops to its lowest theoretical value, absolute zero, the particles' random motion due to their kinetic energy terminates. In the case of DW distribution, the condition of absolute zero temperature or $\theta \to \infty$ translates that there is no common source of disturbance to agitate the text's DWs simultaneously, i.e., each expected DW length state is completely occupied.

Another physically intuitive parameter in Eq. (18) is the parameter $\phi$ in the first exponential term, $e^{\phi}$. This parameter is related to the chemical potential $\mu$ of a grand canonical ensemble of classical particles. In the case of DW distribution, a simple analysis suggests that $\mu = \phi/\theta$ and since $\phi$ has the dimension of energy, as a result the relation between the parameter $\phi$ and the chemical potential energy is $\phi = \mu\beta$. A well-known statement of statistical mechanics is that pressure controls any change in volume and, likewise, chemical potential controls any change in the number of particles. The chemical potential corresponds to the (infinitesimal) change in entropy associated with adding a particle to the system (while holding total energy and volume fixed), $\mu = -T\left(\partial S/\partial N\right)_{U,V}$.

Entropy is another informative thermodynamic property of statistical mechanics systems, and it is defined as the measure of the number of ways in which a system may be arranged; i.e., the measure of disorder. The Boltzmann entropy in statistical mechanics for a system in equilibrium is equal to

$$S = -k_B \sum_i p_i \ln p_i . \tag{24}$$

Therefore, the entropy of the DW distribution in a text can be obtained as

$$S = N k_B \left\{ (\ln N - 1) + \sum_i \frac{n_i}{N} \left[ \ln\left(\frac{\omega^{l_i} l_i^{\alpha}}{n_i}\right) + 1 \right] \right\}. \tag{25}$$

In thermodynamics, the Helmholtz free energy, the thermodynamic potential, is a measure of useful energy or a maximum amount of extractable work from a thermodynamic system, and it is simply defined by

$$F = -k_B T (\ln Z). \tag{26}$$

So by substituting Eq. (20) in Eq. (26), the free energy of the DW organization in a text is equal to

$$F = -k_B T \ln\left(\sum_i \omega^{l_i} l_i^\alpha e^{-\theta l_i}\right) = -\frac{1}{\theta} \ln\left(\sum_i \omega^{l_i} l_i^\alpha e^{-\theta l_i}\right). \tag{27}$$

As we shall see in the following section, these results were put into practice and provided quantitative conclusions for comparative DW organization analysis in terms of thermodynamic concepts.

### 3.3. Some thermodynamic properties of DW organization in selected corpora

In this section, we demonstrate that the SMMA model allows us to obtain and compare the thermodynamic properties of the DW organizations of the previously introduced two corpora. Due to their large sizes, both corpora examined in this study are quite inclusive in terms of the languages' vocabulary (DW) content; for this reason, one can unpretentiously deduce that the thermodynamic properties of the corpora can be extended to the thermodynamic properties of the languages in DW organization as anticipated in the following investigations. Furthermore, for the sake of simplicity, we set Boltzmann's constant $k_B$ equal to 1 for the subsequent computational calculations.

As discussed earlier $\theta = \mathrm{T}^{-1}$; then, the temperature of the corpora, in arbitrary units, was obtained as 0.2264 and 0.2274 for the Brown Corpus and the METU Corpus, respectively. The temperature values indicated that Turkish language is slightly 'hotter' than English language in DW organization. In statistical mechanics terminology, this means that the average energy per DW is somewhat higher in Turkish than in English, which reveals that the DWs are more energetic, or more energy consuming, as used in Turkish. This is an expected result, since the peak of the distribution curve is positioned at longer word length state (~two letters higher) for Turkish, see Figs. (1a) and (1b).

Similarly, the parameter $\phi$ was shown to be analogous to the chemical potential energy given by $\phi = \mu \mathrm{T}^{-1}$, as discussed in the previous section. Hence, the chemical potential energy, in arbitrary units, was calculated as 0.2101 and -0.0653 for the Brown Corpus and the METU Corpus, respectively. These numerical results suggested that the entropy of Turkish language has a tendency to increase with the addition of new DWs, while the entropy of English language has a tendency to decrease with the addition of new DWs. Since the chemical potential energy concept is notoriously somewhat elusive, the full interpretation of the above numerical values in their organizations would require further elaboration.

From Eq. (25), the numerical values of entropies, in arbitrary units, are calculated as $1.8 \times 10^6$ and $9.5 \times 10^6$ for the Brown Corpus and the METU Corpus, respectively. This result uncovered that the increase in disorder is about five times higher in Turkish language's DW organization compared to that of English language.

Finally, the free energies of the corpora's DW organizations were obtained by substituting the SMMA model parameter values, seen in Table 2, into Eq. (27). The numerical values of the free energies, in arbitrary units, are calculated as -2.1911 and -2.8095 for the Brown Corpus and the METU Corpus, respectively. These are the amount of energies have to be committed by language users of two languages in the DW usage during their communication. According to this result, we quantified that the energy consumption due to the DW usage is about 22% less in English compared to Turkish, which suggests that English is more effective language than Turkish at the level of DW organization. This drawn conclusion is in line with the comparison of the average DW length values, 7.8 for the Brown Corpus and 9.4 for the METU Corpus.

From the above straightforward demonstrations, we inferred that the SMMA model transformation methodology is a powerful tool to draw comparative conclusions on organizations presenting MA law behavior by means of enumerating their thermodynamic properties.

## 4. Conclusion

In this study, we proposed a generalization methodology of the MA model, termed as the SMMA model. The derivation of the four parameter SMMA model was based on the statistical mechanics treatment of DWs in a text. The significance of the model is that it consists of an additional structure-dependent parameter input that converts the remaining three parameters into structure-independent free parameters. We have to emphasize that the additional parameter in the SMMA law is not a free parameter during the fitting process. It is uniquely dictated by the formational nature of the structural organization under investigation, and its value is described prior to fitting process. Thus, it is not reasonable to expect that the four-parameter SMMA law provides better fit compared to the three-parameter MA law as indicated by the Akaike information criterion [29]. In fact, we deliberately showed that the MA law and the SMMA law are the identical distribution models with different sets of parameters.

Moreover, the parameters of the SMMA model are associated with a corresponding physical interpretation that can lead to characterization of an organization's thermodynamic properties. The derivation procedure may suggest that the reason for the MA law behavior's inevitable presence in many natural and artificial organizations might be the discrete and energy-preserving nature of such constructs' constituent configuration.

The DW distribution is, of course, only one linguistic trait of texts. However, we emphasize that the methodology used here can be applied to the comparative quantification of regularity between other linguistic organizational levels, languages and even natural phenomena that present the MA law behavior. The recipe for carrying out such investigation is simply to describe; (i) the structural degeneracy parameter for the level of organization under investigation, which is not necessarily the same for each level of organization, and (ii) the energy(effort)-associated states, naturally constituent length or size, in the SMMA model.

Moreover, natural languages serve as a readily available model for the investigation of many complex systems, and linguistic-based theories and algorithms are commonly employed to study such complex systems. Therefore, quantitative linguistic studies not only contribute to the science of language, but also simplify the characterization and understanding of the self-organization and evolution processes of many complex systems. In conclusion, constituent diversity dynamics is of broad scientific interest to a range of disciplines from information technologies to bioinformatics, and we anticipate that the utilization of the presented methodology by those researching complex systems could result in some intriguing outcomes.

**Acknowledgments**

We are grateful to A. Algin for helpful discussions and we appreciate the careful proofreading of the manuscript by H. Kreuzer. This work was partially supported by Eskisehir Osmangazi University's Scientific Research Project Commission (Grant No. 2008-19019).

**TABLES**

**Table 1.** Observed and predicted DW length distributions for the corpora.

| DW length[a] | Number of DWs / Brown Cor. (English) | | | Number of DWs / METU Cor. (Turkish) | | |
|---|---|---|---|---|---|---|
| | Observed[b] | Predicted | | Observed[b] | Predicted | |
| | | MA model[b] | SMMA model | | MA model[b] | SMMA model |
| 1  | 26    | 1     | 1     | 2      | 0      | 0      |
| 2  | 142   | 73    | 73    | 341    | 47     | 47     |
| 3  | 783   | 639   | 640   | 1,592  | 622    | 623    |
| 4  | 2,318 | 2,123 | 2,124 | 3,814  | 2,900  | 2,903  |
| 5  | 4,072 | 4,154 | 4,154 | 9,023  | 7,601  | 7,605  |
| 6  | 5,627 | 5,814 | 5,813 | 13,050 | 13,835 | 13,834 |
| 7  | 6,508 | 6,459 | 6,457 | 19,034 | 19,576 | 19,570 |
| 8  | 6,059 | 6,059 | 6,056 | 22,362 | 23,040 | 23,028 |
| 9  | 5,099 | 4,994 | 4,992 | 23,488 | 23,556 | 23,542 |
| 10 | 3,877 | 3,718 | 3,716 | 22,286 | 21,554 | 21,541 |
| 11 | 2,475 | 2,549 | 2,548 | 18,473 | 18,028 | 18,018 |
| 12 | 1,544 | 1,632 | 1,632 | 14,869 | 14,001 | 13,994 |
| 13 | 885   | 987   | 987   | 10,192 | 10,217 | 10,214 |
| 14 | 444   | 569   | 569   | 6,609  | 7,071  | 7,070  |
| 15 | 209   | 314   | 314   | 3,744  | 4,675  | 4,675  |
| 16 | 85    | 167   | 167   | 2,117  | 2,970  | 2,971  |
| 17 | 48    | 86    | 86    | 1,174  | 1,822  | 1,823  |
| 18 | 23    | 43    | 43    | 607    | 1,084  | 1,085  |
| 19 | 7     | 21    | 21    | 345    | 627    | 628    |
| 20 | 1     | 10    | 10    | 151    | 354    | 354    |
| 21 | 2     | 5     | 5     | 71     | 195    | 196    |
| 22 | 1     | 2     | 2     | 29     | 106    | 106    |
| 23 | NA    | NA    | NA    | 7      | 56     | 56     |
| 24 | NA    | NA    | NA    | 6      | 29     | 29     |
| 25 | NA    | NA    | NA    | 3      | 15     | 15     |

[a] Observed DW length count truncation was 22 letters for the Brown Corpus and and 25 letters for the METU Corpus.
[b] Adopted from Reference [17].

**Table 2.** The fitting analysis results of the MA and the SMMA models for the corpora.

| Model | Corpus | Fitting parameter values[a] and correlation coefficients | | | | |
| --- | --- | --- | --- | --- | --- | --- |
| | | $A\ (\phi)$ | $b\ (\alpha)$ | $c\ (\theta)$ | $R$ | $R^2$ |
| MA model[b] | Brown (English) | $2.5236 \pm 0.4705$ | $8.2039 \pm 0.1864$ | $1.1595 \pm 0.0255$ | 0.9991 | 0.9982 |
| | METU (Turkish) | $0.7454 \pm 0.2858$ | $8.9357 \pm 0.3200$ | $1.0303 \pm 0.0359$ | 0.9973 | 0.9945 |
| SMMA model[c] | Brown (English) | $0.9281 \pm 0.1873$ | $8.2014 \pm 0.1886$ | $4.4173 \pm 0.0262$ | 0.9991 | 0.9982 |
| | METU (Turkish) | $-0.2871 \pm 0.3864$ | $8.9299 \pm 0.3248$ | $4.3970 \pm 0.0371$ | 0.9973 | 0.9945 |

[a] Each value is displayed with the associated standard error.
[b] The free model parameters are $A$, $b$ and $c$.
[c] The free model parameters are $\phi$, $\alpha$ and $\theta$ ($\omega$=26 for the Brown Corpus and $\omega$=29 for the METU Corpus).

**FIGURES**

**Figure 1.** The prediction of the DW distribution for the Brown and the METU Corpora. Observed number of DW versus DW length data (dashed line) are fitted with distribution prediction curves (solid lines) proposed by the MA model and the SMAA model for (a) the Brown Corpus (English) and (b) the METU Corpus (Turkish). Note that both prediction curves, obtained by the MA model and the derived SMMA model, exactly overlap for both distributions.

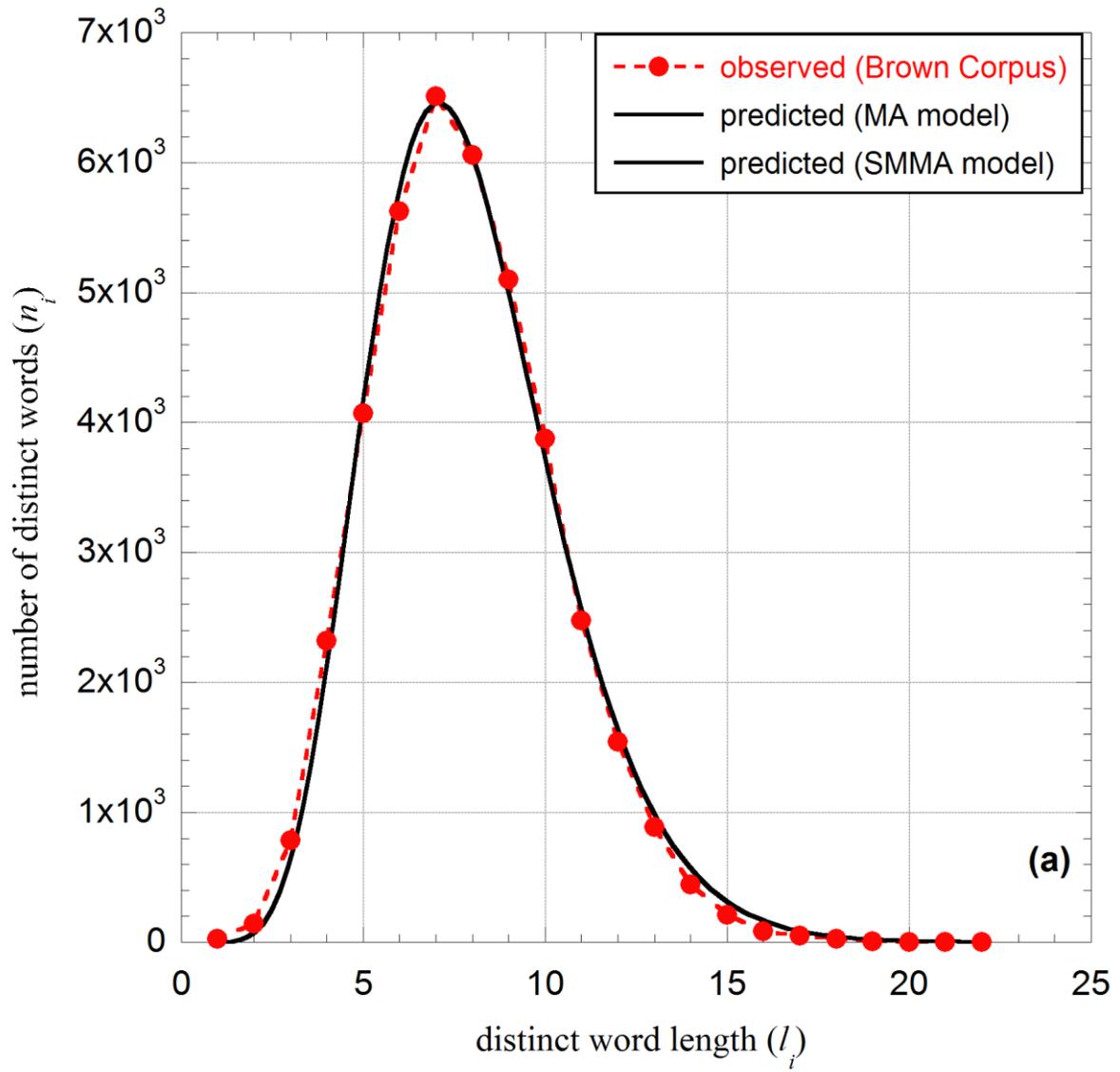

**Fig. 1a**

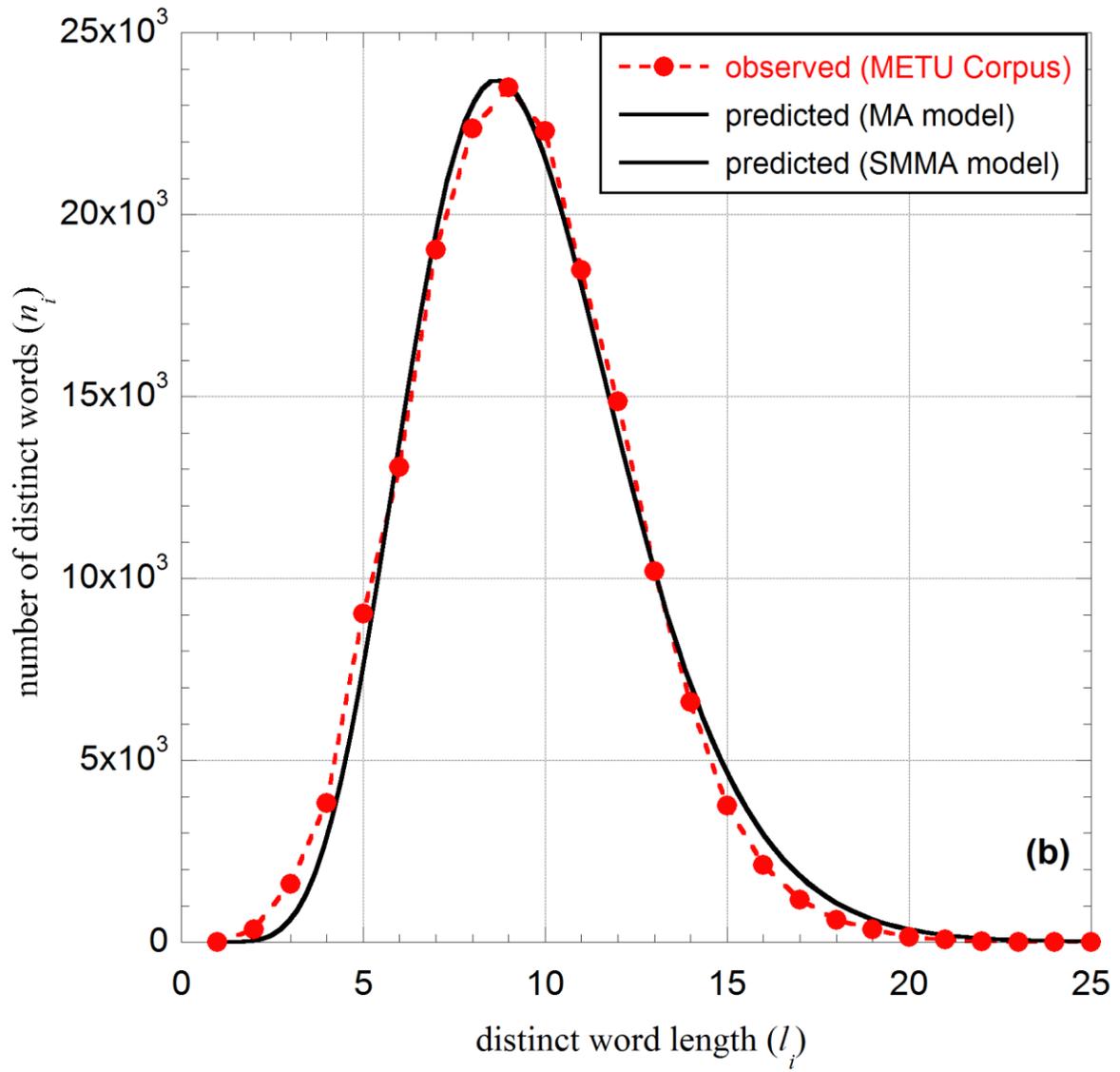

**Fig. 1b**